# Superconductivity in WBe$_2$

J. S. Kim,[1] P. M. Dee,[1-3] J. J. Hamlin,[1] P. J. Hirschfeld,[1] and G. R. Stewart[1]

[1]Department of Physics, University of Florida, Gainesville FL 32611, USA

[2]Department of Materials Science and Engineering, University of Florida, Gainesville, FL 32611, USA

[3]Oak Ridge National Laboratory, Oak Ridge TN 37830, USA

Corresponding author: G R Stewart, gstewart@ufl.edu, orcid 0000-0001-7251-1684

Statements and declarations: The authors have no competing interests to declare.

Acknowledgements: Work at the University of Florida was performed under the auspices of the U.S. National Science Foundation, Division of Materials Research under Contract No. NSF-DMR-2118718.

**Abstract:** WBe$_2$, which occurs in space group 194, with hexagonal symmetry P6$_3$/mmc, is prepared by arc-melting at temperatures above 2200 C, where Be vapor loss is significant. This study is motivated by recent work on MoB$_2$ and WB$_2$, both superconductors (T$_c$=32 and 17 K respectively) under high (~70 GPa) pressure. In order to avoid the known Be-rich superconducting phases (WBe$_{13}$ and WBe$_{22}$) in the complex phase diagram, both known to be superconducting at 4.1 K, the sample was prepared with a slight

(~5%) excess of W. The resultant sample, prepared using high purity (99.999%) Be, is essentially single phase $WBe_2$, with some spread in its superconducting properties due to the known homogeneity range. ($WBe_2$ forms in space group 194 between approximately $W_{1.02}Be_{1.98}$ and $W_{0.88}Be_{2.12}$.) Characterization was carried out with x-ray diffraction, electrical resistivity, $\rho$, in zero and applied magnetic fields, and specific heat. The resistivity in zero and applied fields and specific heat data indicate that our sample of $WBe_2$ is a bulk superconductor at ambient pressure with a $T_c^{onset}$ in $\rho$ at 1.05 K, and $T_c(\rho \rightarrow 0)$ at ~0.86 K. There is no signature of superconductivity in $\rho$ at 4.1 K, indicating successful avoidance of $WBe_{13}$ and $WBe_{22}$. The $\rho$ data in field indicate an upper critical field of approximately 400 gauss.

1. Introduction

Recently, superconductivity at 32 K under high pressure was found [1] in $MoB_2$ in the hP3, space group 191 structure – the same structure as the 39 K ambient pressure superconductor $MgB_2$. Our follow up work [2] found that superconductivity occurs in $WB_2$, space group 194, at 17 K and starting at pressures above 50 GPa. This result was understood via theoretical calculations that show that electron-phonon mediated superconductivity in $WB_2$ under pressure originates from the formation of metastable stacking faults and twin boundaries that exhibit a local structure resembling $MgB_2$ (hP3, space group 191, prototype $AlB_2$).

Since our work on $WB_2$, we have investigated [3] various binary compounds occurring in space group 194. The current work discusses the preparation and characterization at ambient pressure of $WBe_2$, which also occurs in this space group, although with different atomic Wyckoff positions. The authors in ref. 4, in addition to finding no superconductivity in $WBe_2$ down to their lowest temperature of measurement (1.68 K) also reported the same lack of superconductivity in MBe$_2$, M=Re, Mo, Zr.

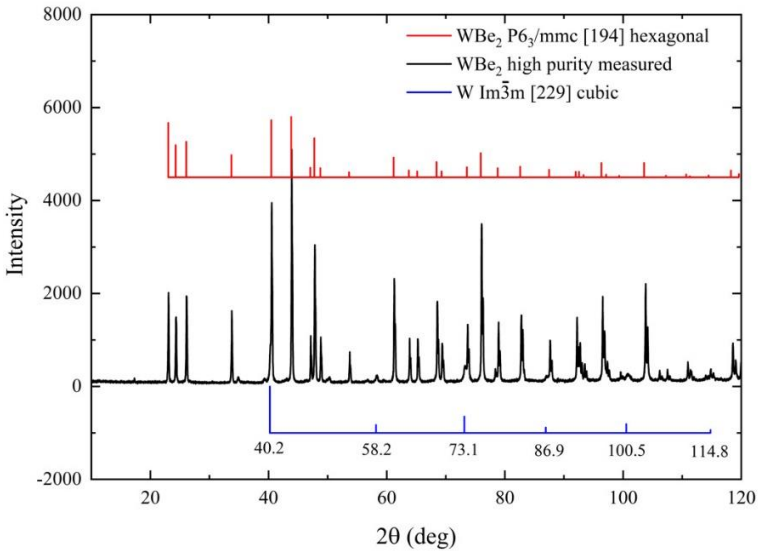

2. Sample Preparation

Pure W metal foil (0.1 mm thick), with purity 99.95% from Alfa Aesar, was mixed together with pieces of Be, purity 99.999% from Atomergic, and then arc-melted. Due to the extremely high vapor pressure (~ 100 mm Hg) of Be at the melting point of WBe$_2$ (T$_m$ ~ 2200 C), a large excess (~30 %) of Be was used while melting the components together a total of six times (i. e. according to our experience, there was approximately a 5% loss of Be during each melt). The x-ray diffraction pattern of a powdered piece of the resulting arc-melted button is shown in Fig. 1.

Fig. 1 X-ray diffraction pattern of WBe$_2$ (measured, black line), of the Materials Project pattern for WBe$_2$ (red line), and of pure W (blue line.) (The diffraction patterns from the Materials Project, calculated by DFT, are essentially the same as those found in the cards [18-240, WBe$_2$; 4-0806, W] found in the Joint Committee on Powder Diffraction Standards [JCPDS] book for x-ray diffraction, except that the JCPDS pattern for WBe$_2$ only extends up to 80 degrees 2θ.) The W diffraction lines at 2Θ = 40.2 and 72 overlie intense WBe$_2$ lines, but the W lines near 58.2 and 100.5 do not overlie the WBe$_2$ pattern. Thus, from the intensities in the measured pattern there appears to be some trace amount (~5-10%) of W metal in the sample. As will be discussed below, there is no resistivity evidence for either WBe$_{22}$ or WBe$_{13}$ (which both exhibit [4-6] superconductivity around 4.1 K).

This is consistent with the measured x-ray pattern since both of these higher beryllide compounds have strong peaks in their x-ray diffraction patterns below (or just above) 20 degrees 2θ of which there is *no* evidence in our diffraction pattern. Thus, we estimate that our sample has less than 2% of $WBe_{13}$ or $WBe_{22}$, and is approximately 90-95% pure $WBe_2$ with lattice parameters a= 4.452 Å c = 7.309 Å (vs Materials Project lattice parameters of a=4.44 Å and c=7.31 Å and the JCPDS values of 4.45 Å and 7.30 Å). Our measured lattice parameters are consistent with the sample being on the W-rich side of the homogeneity range for $WBe_2$, see ref. [7] for the lattice parameters for the W- and Be-rich compositions in the $WBe_2$ range of stability. This is also consistent with the x-ray data in Fig. 1.

### 3. Characterization by:
#### 3.1 Resistivity:

The low temperature resistivity in zero magnetic field, measured using standard 4 wire DC measurement on a thin (~0.5 mm thick) slice of our $WBe_2$, is shown in Fig. 2. The resistivity data in applied field are shown in Fig. 3. As discussed in the previous section, there is no indication of the intense x-ray diffraction lines near 20 degrees 2θ for $WBe_{13}$ (at 2θ ~ 22 degrees) or for $WBe_{22}$ (at 2θ ~ 13 degrees). As seen in the resistivity data, there is also no indication of the $T_c$ of either $WBe_{13}$ or $WBe_{22}$ (both have [4-6] $T_c$ = 4.1 K). There is instead a sharp transition down to ρ=0 starting at 1.05 K and ending with ρ→0 below 0.86 K. (In our measurement method, 'zero' resistivity corresponds to ρ values below 0.86 K between 0.006 and 0.4 μΩ-cm.) Published work [4] on $WBe_2$ previously indicated no superconductivity down to a lowest temperature of measurement of 1.68 K. Thus, these resistivity data indicate that $WBe_2$ is a superconductor itself, in addition to the other known 1:13 and 1:22 stoichiometries.

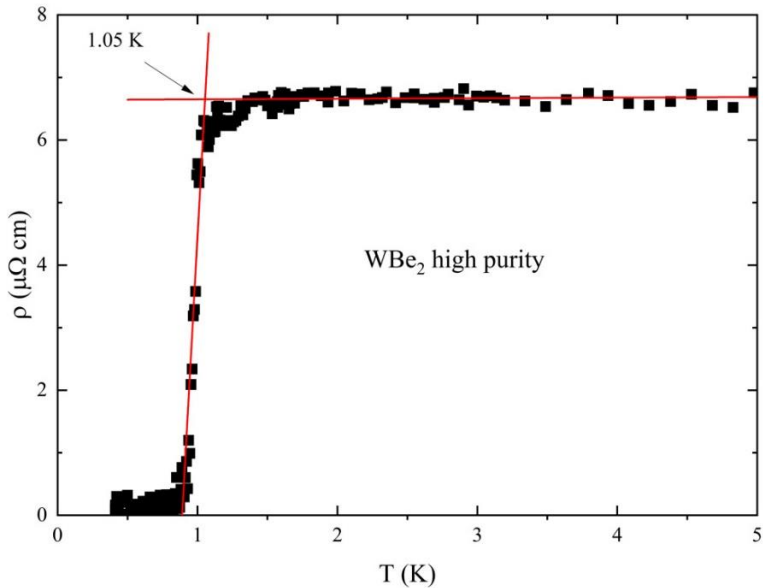

Fig. 2 Resistivity of our WBe$_2$ sample between 0.3 and 5 K, measured using a four point DC method. Due to the small size of the sample, there is about ±5% noise in the data. Due to uncertainties in the precise geometrical size of the sample, the absolute error bar is ~10%. The residual resistivity ratio (RRR), equal to the ratio of the room temperature resistivity to the resistivity in the normal state extrapolated to T=0 (R(300 K)/ R$^{normal}$(T→0)) is 8.23 (which, since it involves a ratio, is of course not affected by the geometrical uncertainty). No indication of superconductivity at T$_c$ = 4.1 K is present. More interesting is the relatively sharp transition down towards ρ=0 starting at 1.05 K. Field data are shown below in Fig. 3.

**Fig. 3** Resistivity of high purity WBe$_2$ in applied fields. In the inset, the critical temperatures vs field are plotted to arrive at the zero temperature upper critical field of approximately 0.04 T, four times larger than elemental Al, which also has T$_c$ ~ 1 K.

### 3.2 Specific Heat:

Turning now to a bulk measure of superconductivity, the specific heat of WBe$_2$, measured in an apparatus using the time constant method as described in ref. [8], is shown in Fig. 4. The resistive transition in zero field shown in Fig. 2 between 0.86 and 1.05 K corresponds to a bulk superconducting transition in the whole sample as measured by the specific heat, with an onset temperature of ~0.88 K.

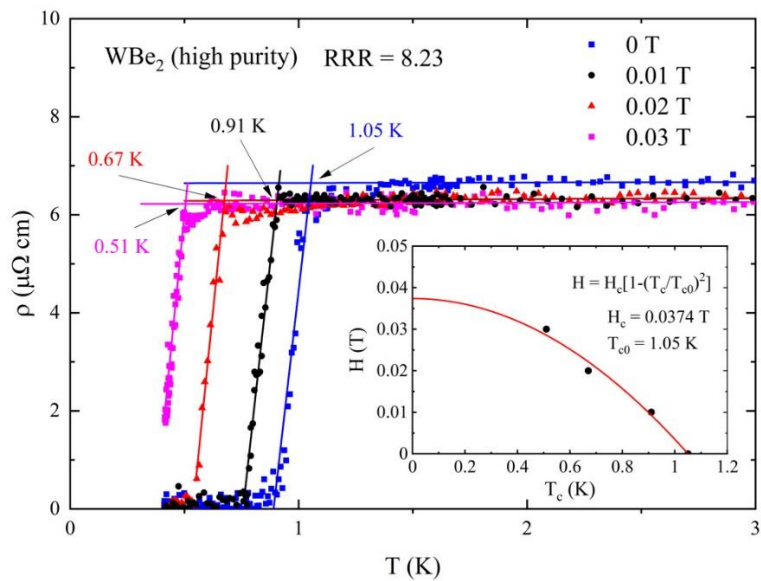

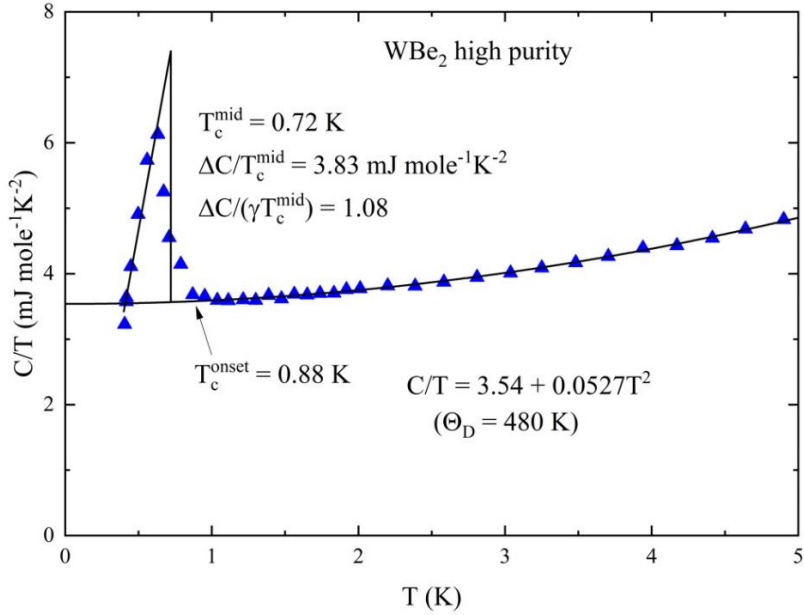

Fig. 4 Specific heat, C, divided by temperature vs temperature of the same piece of $WBe_2$ used for the resistivity measurement (data in Figs. 2 and 3. Absolute error bars for the data (not shown) are approximately ±3% [8]. The peak in the specific heat transition occurs at ~0.7 K, consistent with the zero field resistivity data in Fig. 2. The specific heat γ value, 3.54 mJ/molK², (where at low temperatures in the normal state $C/T = γ + βT^2$) is proportional to the bare density of states N(0) via N(0)(1+λ) = 0.4244γ. The electron-phonon coupling parameter λ can be estimated from the McMillan formula [9] (using $T_c$=1.05 K, the Debye temperature = 480 K, and μ*=0.13) to be ~0.44. This results in a bare density of states, N(0), of 1.04 states/eV-atom.

## 4. Conclusions and Summary

While studying binary compounds with crystal structure in space group 194, $P6_3$/mmc, in a follow up to our work [2] with $WB_2$, we have discovered bulk superconductivity in $WBe_2$ at 1 K.

It is interesting to speculate why the $T_c$ in the newly discovered hexagonal superconductor $WBe_2$ (1 K) is so much smaller than those found in cubic $WBe_{13}$ and $WBe_{22}$ (both $T_c$ values are [4-6] around 4.1 K.) One possibility is the difference in structure in the compounds. $WBe_{13}$ occurs in a cage structure, the same as all of the $MBe_{13}$ compounds including the heavy Fermion $UBe_{13}$, with the shortest W-Be distance equal to 2.19 Å while in $WBe_{22}$ each W is also surrounded by Be, bonded to 16 Be atoms, with W-Be distances varying between 2.51 and 2.53 Å. In contrast, in $WBe_2$ the W atom is bonded to 12 Be atoms in a more open structure, all at the larger distance of 2.61 Å.

This difference in structure will affect the fundamental parameters (N(0) and λ) in the standard BCS electron phonon coupling model ($T_c \propto$ exp(-1/N(0))) where a larger electronic density of states leads to higher $T_c$. For example, consider $WBe_{22}$ for which specific heat data exist [5], γ=10.0 mJ/molK^2, $\Theta_D$ =960 K, and λ = 0.47 from the McMillan formula gives a density of states per W-atom, N(0), almost three times that found in $WBe_2$.

A phenomenological model for $T_c$ ($T_c$ = ($\Theta_D$/20)*( λ – 0.25)) proposed [10] by John Rowell, which focuses more on the lattice stiffness ($\propto \Theta_D$), also is qualitatively consistent with a higher $T_c$ for $WBe_{22}$ with its much higher Debye temperature due to the cage structure of the Be atoms.

Work is underway to characterize $WBe_2$, both theoretically and experimentally, as a function of high pressure.